\documentclass[]{aa}

\usepackage{natbib}
\bibpunct{(}{)}{;}{a}{}{,}

\usepackage{graphicx}

\usepackage{txfonts}

\begin{document}

   \title{Clouds in the atmospheres of extrasolar planets}

   \subtitle{I. Climatic effects of multi-layered clouds for Earth-like planets and implications for habitable zones}

   \author{D. Kitzmann
          \inst{1},
          A.B.C. Patzer
          \inst{1},
          P. von Paris
          \inst{2},
          M. Godolt
          \inst{1},
          B. Stracke
          \inst{2},
          S. Gebauer
          \inst{1},
          J.L. Grenfell
          \inst{1},
          \and
          H. Rauer
          \inst{1,2}
          }
          
   \authorrunning{D. Kitzmann et al.}
   \titlerunning{Clouds in the atmospheres of extrasolar planets. I.}

   \offprints{D. Kitzmann}

   \institute{Zentrum f\"ur Astronomie und Astrophysik, Technische Universit\"at Berlin,
              Hardenbergstr. 36, 10623 Berlin (Germany)\\
              \email{kitzmann@astro.physik.tu-berlin.de}
              \and
              Institut f\"ur Planetenforschung, Deutsches Zentrum f\"ur Luft- und Raumfahrt (DLR),
              Rutherfordstr. 2, 12489 Berlin (Germany)
             }

   \date{Received 19 October 2009 / Accepted 22 December 2009}

  \abstract
   {}
   {The effects of multi-layered clouds in the atmospheres of Earth-like planets orbiting different types of stars are studied. The radiative effects of cloud particles are directly correlated with their wavelength-dependent optical properties. Therefore the incident stellar spectra may play an important role for the climatic effect of clouds. We discuss the influence of clouds with mean properties measured in the Earth's atmosphere on the surface temperatures and Bond albedos of Earth-like planets orbiting different types of main sequence dwarf stars. The influence of clouds on the position of the habitable zone around these central star types is discussed.}
   {A parametric cloud model has been developed based on observations in the Earth's atmosphere. The corresponding optical properties of the cloud particles are calculated with the Mie theory accounting for shape effects of ice particles by the equivalent sphere method. The parametric cloud model is linked with a one-dimensional radiative-convective climate model to study the effect of clouds on the surface temperature and the Bond albedo of Earth-like planets in dependence of the type of central star.}
   {The albedo effect of the low-level clouds depends only weakly on the incident stellar spectra because the optical properties remain almost constant in the wavelength range of the maximum of the incident stellar radiation. The greenhouse effect of the high-level clouds on the other hand depends on the temperature of the lower atmosphere, which is itself an indirect consequence of the different types of central stars. In general the planetary Bond albedo increases with the cloud cover of either cloud type. An anomaly was found for the K and M-type stars however, resulting in a decreasing Bond albedo with increasing cloud cover for certain atmospheric conditions. Depending on the cloud properties, the position of the habitable zone can be located either farther from or closer to the central star. As a rule, low-level water clouds lead to a decrease of distance because of their albedo effect, while the high-level ice clouds lead to an increase in distance. The maximum variations are about $15 \%$ decrease and $35 \%$ increase in distance compared to the clear sky case for the same mean Earth surface conditions in each case.}
   {}


   \maketitle


\begin{table*}[t]
  \caption[]{Mean cloud properties obtained from surface and satellite measurements.}
  \label{tabWarren07}
  \centering
  \begin{tabular}{l r r r r r r}
    \hline
    \noalign{\smallskip}
     & \multicolumn{3}{c}{Cloud coverage (\%)$^{\mathrm{a}}$} & \multicolumn{3}{c}{Global average cloud properties$^{\mathrm{b}}$}\\
    Cloud type & Ocean & Land & Global mean & Top temperature & Top pressure & Optical depth$^{\mathrm{c}}$\\
    \noalign{\smallskip}
    \hline
    \noalign{\smallskip}
    \textbf{Low-level clouds} & 47 & 22 & 39.5 & $281.1 \ \mathrm{K}$ & $0.826 \ \mathrm{bar}$ & 4.7\\
    Stratus & 12 & 5 & 9.9 & & &\\
    Stratocumulus & 22 & 12 & 19.0 & & &\\
    Cumulus & 13 & 5 & 10.6 & & &\\
    Cumulonimbus$^{\mathrm{d}}$ & 6 & 4 & 5.4 & & &\\
    \noalign{\smallskip}
    \textbf{High-level clouds} & 12 & 22 & 15.0 & $227.5 \ \mathrm{K}$ & $0.267 \ \mathrm{bar}$ & 2.2\\
    Cirriform & 12 & 22 & 15.0 & & &\\
    \noalign{\smallskip}
    \hline
  \end{tabular}
  \begin{list}{}{}
    \item[$^{\mathrm{a}}$] taken from surface observations \citep{Warren07}, the global mean coverages have been derived by a weighted average from the coverages over ocean ($70\%$) and land ($30\%$)
    \item[$^{\mathrm{b}}$] from satellite observations within the ISCCP (see \citet{Rossow99a} for details)
    \item[$^{\mathrm{c}}$] optical depth at $0.6 \, \mathrm{\mu m}$
    \item[$^{\mathrm{d}}$] not included in the model and in the calculation of the mean coverages
  \end{list}
\end{table*}

\section{Introduction}

Cloud particles can have an important impact on the climate of planetary atmospheres by either scattering the incident stellar radiation back to space (albedo effect) or by trapping the infrared radiation in the atmosphere (greenhouse effect). The answer to the question which of these effects dominates for a given cloud type depends on a variety of cloud parameters, the most important of which are the cloud composition (water, carbon dioxide etc.), the size distribution of the cloud particles, the optical depth of the cloud layer, multi-layered cloud coverage and the cloud altitude.

In the case of the well known Earth atmosphere, where clouds are a very common phenomenon with a mean global cloud coverage of more than $50 \%$, low-level water clouds have a net cooling effect on the surface, while high-level ice clouds exhibit a greenhouse effect, resulting in surface heating. A comprehensive review about the climatic effects of clouds in the Earth atmosphere can be found in e.g. \citet{KondratyevCloud} and references therein. The albedo and the greenhouse effect are directly correlated with the wavelength-dependent optical properties of the cloud particles. The incident stellar spectra in combination with these optical properties of the cloud may therefore play an important role for the surface temperature.

Model calculations regarding the habitability of planets orbiting different types of central stars have already been done by e.g. \citet{Kasting1993},
\citet{Segura03, Segura05} or \citet{Selsis2007}. Still, these models lack a detailed treatment of the cloud radiative forcing in the radiative transfer and aim only to mimic the cooling effect of low-level clouds by an adjusted surface albedo. Focussing on the inner boundary of the habitable zone around the Sun \citet{Kasting1988} included the effect of one very extended water droplet cloud with rather simplified optical properties in some of the model scenarios. Without taking clouds explicitly into account in their atmospheric climate model calculations \citet{Kaltenegger2007} studied the emission spectra of Earth-like planets at different evolutionary stages.

We study the effects of different incident stellar spectra in conjunction with multi-layered clouds of different types on the surface temperatures of Earth-like extrasolar planets, which determine the potential habitability of a terrestrial planet.
As a first approximation, a simplified cloud description scheme using parametrised size distribution functions is used. Other cloud properties like optical depth and cloud top pressure have been taken from measurements. The parametric cloud model is described in Sect. \ref{secCloudDescription}.
To study the basic climatic effects, our cloud scheme has been coupled with a one-dimensional radiative-convective climate model which includes the possibility to account for different amounts of cloud coverages as well as the partial overlap of two cloud layers, as described in Sect. \ref{secClimateModel}. In Sect. \ref{secEarthReference} we apply this climate model including the parametrised cloud description to the modern Earth atmosphere to verify the applicability of our model approach.
Resulting surface temperatures and Bond albedos for different cloud types in the atmospheres of Earth-like planets orbiting different types of central stars and implications for the positions of the habitable zones are presented in Sect. \ref{secEarthLikeAtmos}.


\section{Cloud model description}
\label{secCloudDescription}

Based on the extremely well-studied properties of different cloud types occurring in the Earth's atmosphere we developed a (parametrised) multi-layered cloud description scheme. Two different kinds of cloud layers are considered here: low-level water droplet and high-level water ice clouds. The corresponding global and temporal average cloud coverages resulting from long-term surface observations have been published by e.g. \citet{Warren07} and are given in Table \ref{tabWarren07}. The measured cloud coverages already include a certain amount of overlap between the different cloud layers, which was not further specified by \citet{Warren07}. Average properties of Earth clouds, for instance the cloud top temperature and pressure as well as their optical depth have been derived by long-term satellite-based measurements within the International Satellite Cloud Climatology Project (ISCCP) for example. Global and temporal mean cloud properties based on these surveys have been published by \citet{Rossow99a} and are also summarised in Table \ref{tabWarren07}.

Other cloud types present are not taken into account. With a mean coverage of $17 \%$ the most common mid-level clouds, altocumulus clouds, have been reported to be radiatively neutral, which means that their albedo and greenhouse effect balance each other \citep{Poetzsch95}, which justifies our approach to neglect them. Cumulonimbus clouds are also excluded from our cloud description scheme. These clouds can extend up to $10 \ \mathrm{km}$, which makes it difficult to include them in our one-dimensional climate model (see Sect. \ref{secClimateModel}). For such extensive vertical clouds not only the radiation entering the cloud from the top or bottom has to be considered, but also photons entering the cloud sideways have to be taken into account. This though is essentially a three-dimensional problem which cannot be treated by a one-dimensional model. Since such clouds can not be properly included and also present only $5 \%$ of the global cloud coverage, we decided against implementing them into our column model. 

Apart from these measured properties the size distribution of the cloud particles has also to be known to calculate the wavelength-dependent absorption and scattering properties of a cloud. Usually it is not appropriate to describe a cloud with uniform-size particles (but see e.g. \citet{Mitchell02} and \citet{McFarquhar98} concerning the applicability of an effective radius). The radius $a$ of a cloud particle has therefore to be treated as a random variable, characterised by a distribution function $f(a)$. Normally these distributions have to be calculated from first principles accounting for all relevant microphysical cloud processes (see \citet{PruppKlettMicro}), which involves a great deal of computation. For the Earth's atmosphere it is possible though to derive parametric analytical distribution functions based on measurements. Since the focus of this work is on Earth-like planetary atmospheres, we assume that these analytical distribution functions are also valid for the atmospheres considered here. This assumption neglects any possible influence of different atmospheric properties (like differences in the atmospheric dynamics or chemical composition) on the distribution function of the cloud particles. Such differences could arise when considering planets with for instance different rotation periods or a different landmass distribution compared to Earth. Therefore the cloud parametrisations used here are limited to Earth-like planets.

\subsection{Low-level cloud particle size distribution}

Observations of low-level clouds in the Earth's atmosphere show that the measured size distribution of the cloud particles can be well-represented by a log-normal distribution
\begin{equation}
  f(a) = n \frac{1}{\sqrt{2\pi} \sigma a} \exp \left( - \frac{ \left( \ln a - \ln a_m \right)^2}{2 \sigma^2} \right) \quad .
\end{equation}
Measured parameters for the particle density $n$, the mean particle radius $a_m$ and the standard deviation $\sigma$ can be found for instance in \citet{KokhanovskyCloud}. The mean values for continental and maritime water clouds are given in Table \ref{tab2}.

\begin{table}[t]
  \caption[]{Parameter sets for the description of maritime and continental water clouds by a log-normal distribution.}
  \label{tab2}
  \centering
  \begin{tabular}{l c c}
    \hline
    \noalign{\smallskip}
    Parameter & Maritime clouds & Continental clouds\\
    \hline
    \noalign{\smallskip}
    $n \ (\mathrm{cm^{-3}})$ & 91.0 & 254.0 \\
    $a_m \ (\mathrm{K})$ & 6.0 & 4.0 \\
    $\sigma$ & 0.4 & 0.4 \\
    \hline
    \noalign{\smallskip}
  \end{tabular}
\end{table}

The distribution functions using these sets of parameters are averaged according to the total ocean ($70\%$) and land ($30\%$) surface area of the Earth. The resulting mean distribution function is used to represent low-level clouds in our climate model. We also assume that all low-level cloud droplets are spherical particles composed of pure liquid water. The optical properties from the obtained distribution of the cloud particles can then be calculated with the Mie theory (see Sect. \ref{subMieTheory}).

\subsection{Size distributions of high-level cloud particles}

High-level clouds represent a much more complex system, because the ice crystals can have a huge variety of shapes, which makes the description by a single distribution function and the calculation of the optical properties much more complicated.
The most common shapes present in these clouds are solid and hollow columns, plates, bullets and bullet rosettes. For simplicity all ice particles are considered as solid hexagonal columns throughout this work.

Based on in-situ measurements in cirrus clouds, \citet{Heymsfield84} derived analytical size distributions for high-level ice clouds using a power law size distribution 
\begin{equation}
  f(a_c) = A a^B_c \quad ,
  \label{Heymsfield_distribution}
\end{equation}
where $A$ is the intercept parameter, $B$ the slope, and $a_c$ the column's maximum dimension.
Depending on the cloud's temperature, a bimodal or unimodal distribution must be used. For the temperature range in our case (see Table \ref{tabWarren07}) a unimodal distribution is sufficient to describe the measured size distribution. The parameters $A$ and $B$ are temperature dependent and for the temperature range in question given by \citep{Heymsfield84}
\begin{eqnarray}
  A & = & \frac{2.55 \cdot 10^{-5}}{100^B} \ \mathrm{cm^{-3}} \mathrm{\mu m^{-1}} \\
  B & = & -3.23 \quad .
\end{eqnarray}

Since the size distribution derived by \citet{Heymsfield84} depends only on the maximum particle dimension, we additionally need to prescribe the aspect ratio of the particles to fully describe the columns. \citet{Heymsfield72} provided analytical expressions for the aspect ratio $\Gamma$ of the column base width $D$ and the column length $L$ valid for temperatures near $\approx -20 \ \mathrm{^\circ C}$. For $L \leq 200 \ \mathrm{\mu m}$ the aspect ratio is given by
\begin{equation}
  \Gamma(L) = 2 \ ,
\end{equation}
whereas
\begin{equation}
  \Gamma(L) = 5 \cdot L^{0.59}
\end{equation}
is used for crystals with $L > 200 \ \mathrm{\mu m}$ .

Because the length $L$ is always longer than the base width $D$, for columns with these aspect ratios the maximum dimension $a_c$, which is used to parametrise the distribution function (Eq. \ref{Heymsfield_distribution}), refers to the crystal's length in all cases.

Even assuming exclusively solid hexagonal columns as particle shapes, the calculation of the corresponding optical properties is nevertheless more complex than in the case of spherical droplets because the Mie theory cannot be applied directly. There do exist some applications for the derivation of the optical properties of these non-spherical particles, but they are either limited in their validity range (e.g. geometrical optics or ray tracing methods) or need an excessive amount of computing time (e.g. finite-time-domain theory).

However, it is possible to introduce so-called \textit{equivalent spheres}, i.e. the non-spheric particles are replaced by spheres, which mimic their optical properties. In this way the Mie theory can be used again allowing a fast computation.
Commonly used are equal-volume spheres and equal-area spheres, whereby a sphere has the same volume or the same surface area as the respective non-spherical particle. In both cases the number density of the equivalent spheres and the non-spherical particles remain the same, while either conserving the total volume (in the equal-volume sphere case) or the total surface area (for the equal-area sphere case) of the non-spherical particles. Still, as pointed out by \citet{Grenfell99}, these approaches yield mostly too small scattering albedos and too large asymmetry parameters in comparison with the non-spherical particles. A much better agreement is achieved by using spheres having the same volume-to-surface ratio as the non-spherical particles \citep{Grenfell99}. But to conserve the total volume and area of the non-spherical particles, the number density of the equivalent spheres has to be adapted. Compared to the equal-volume or area spheres, the sizes resulting from the volume-to-surface equivalent sphere method are generally smaller, which in turn leads to smaller asymmetry parameters and larger scattering albedos. This implies that the volume-to-surface equivalent sphere approach offers a much better approximation for the calculation of the optical properties of the non-spherical particles in most cases.

The application of the volume-to-surface sphere approach for hexagonal columns was published by \citet{Neshyba03}. Following their treatment the radius of the equivalent spheres for hexagonal columns is given by the expression
\begin{equation}
  a = a_c \frac{3 \Gamma \sqrt{3} }{4 \Gamma + \sqrt{3}} \ ,
\end{equation}
while the number density of the spheres $n_s$ is determined by
\begin{equation}
  n_s = n \, \frac{\sqrt{3}}{12 \pi \; \Gamma^2} \ .
\end{equation}

As noted by \citet{Neshyba03} these approximations have only been tested for use in energy budget studies, i.e. when using angle-averaged properties of the radiation field (e.g. the spectral flux) as done in this study (cf. \ref{subSec_RadiativeTransfer}). Their applicability might though be limited concerning the calculation of angle-dependent spectral intensities. In particular the required assumption of randomly oriented columns is questionable, because most of the columns are falling with their long axes parallel to the ground as confirmed by observations of e.g. \citet{Ono69}.

\subsection{Optical properties of cloud particles}
\label{subMieTheory}

For spherical particles within the size interval $(a + da)$ the usual transport coefficients are given by
\begin{eqnarray}
  \chi_\lambda(a) da & = & \pi a^2 Q_{\lambda}^\mathrm{ext} f(a) da\\
  \overline{s}_\lambda(a) da & = & \pi a^2 Q_{\lambda}^\mathrm{sca} f(a) da\\
  \kappa_\lambda(a) da & = & \pi a^2 Q_{\lambda}^\mathrm{abs} f(a) da 
                         = \chi_\lambda(a) da - \overline{s}_\lambda(a) da \ ,
\end{eqnarray}
where $\chi_\lambda$ denotes the extinction coefficient, $\overline{s}_\lambda$ the (total) scattering coefficient and $\kappa_\lambda$ the absorption coefficient.
Integration over the whole size range yields the complete contribution from the distribution function $f(a)$ to the total values of the optical properties
\begin{eqnarray}
  \chi_\lambda & = & \int_0^\infty \chi_\lambda(a) da\\
  \overline{s}_\lambda & = & \int_0^\infty \overline{s}_\lambda(a) da\\
  \kappa_\lambda & = & \int_0^\infty \kappa_\lambda(a) da \ .
\end{eqnarray}

The required optical properties (extinction efficiency $Q_{\lambda}^\mathrm{ext}$, scattering efficiency $Q_{\lambda}^\mathrm{sca}$, and absorption efficiency $Q_{\lambda}^\mathrm{abs}$) are calculated here using the Mie theory \citep{BohrenHuffmanScat}. Since the Mie calculations become computationally prohibitive for large size parameters, the large particle limit is employed in these cases.
Using the large particle limit the efficiencies are given by the simple expressions
\begin{eqnarray}
  Q_{\lambda}^\mathrm{ext} & = & 2\\ 
  Q_{\lambda}^\mathrm{sca} & = & 1 + Q_{\lambda}^\mathrm{refl}\\
  Q_{\lambda}^\mathrm{abs} & = & 1 - Q_{\lambda}^\mathrm{refl} \ ,
\end{eqnarray}
where the reflection efficiency $Q_{\lambda}^\mathrm{refl}$ can be calculated with the geometric optics approximation which is valid for particles much larger than the considered wavelength.

Instead of the full scattering phase function $p_\lambda(\theta,a)$, the asymmetry parameter $g_\lambda(a)$, which is defined by
\begin{equation}
  g_\lambda(a) = \frac{1}{2} \int_0^\pi p_\lambda (\theta,a) \cos \theta \sin \theta d\theta
\end{equation}
for a single particle of radius $a$ and the scattering angle $\theta$ can be used to approximate the effects of multiple scattering.
The asymmetry parameter can be written as a function of the Mie coefficients and the size parameter \citep[see][]{BohrenHuffmanScat}.

For a continuous distribution function as in the cloud description, the average value
\begin{equation}
  g_\lambda = \frac{ \int_0^\infty \pi a^2 g_\lambda(a) Q_{\lambda}^\mathrm{sca} f(a) da}
                   { \int_0^\infty \pi a^2 Q_{\lambda}^\mathrm{sca} f(a) da}
\end{equation}
is used, applicable for multiple scattering.

In the case of the low-level clouds we use the refractive indices for pure liquid water taken from \citet{Segelstein81}, whereas for the high-level clouds the indices for water ice, published by \citet{Warren08} are adopted. The refractive indices are assumed to be independent of temperature.

\begin{figure}
  \centering
  \resizebox{\hsize}{!}{\includegraphics{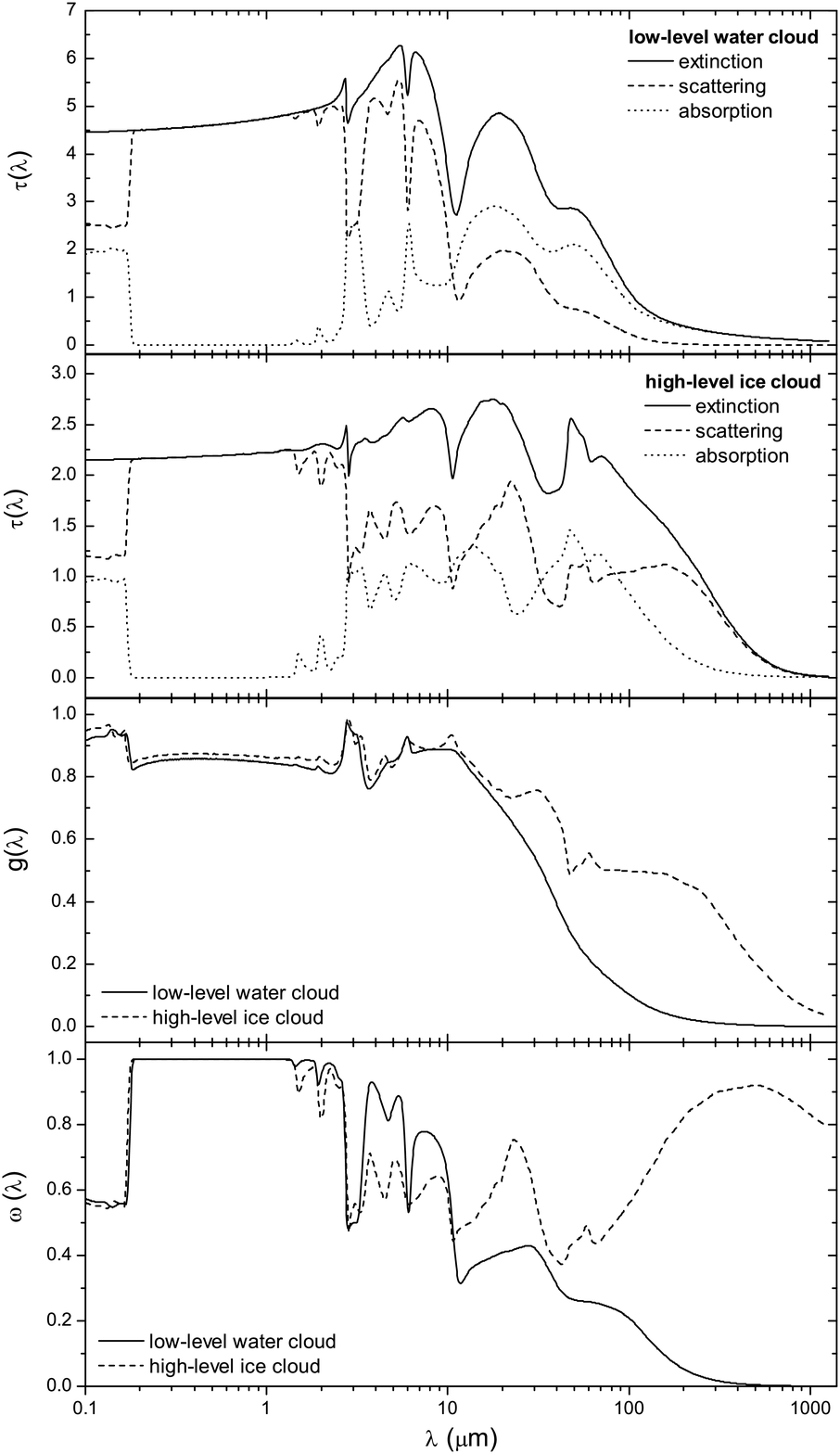}}
  \caption{Calculated optical properties of the cloud model. \textbf{Upper diagram:} optical depth of the low-level water cloud (solid line) with the individual contributions of absorption (dotted line) and scattering (dashed line); \textbf{upper middle diagram:} optical depth of the high-level ice cloud (solid line) with the individual contributions of absorption (dotted line) and scattering (dashed line); \textbf{lower middle diagram:} asymmetry parameter $g_\lambda$ of the water cloud (solid line) and ice cloud (dashed line); \textbf{lower diagram:} scattering albedo $\omega_\lambda$ of the water cloud (solid line) and ice cloud (dashed line).}
  \label{opticaldepth}
\end{figure}

The results of the Mie theory calculations for high- and low-level cloud size distributions have been scaled according to the measured optical depth from Table \ref{tabWarren07} and are shown in Fig. \ref{opticaldepth}.
Scattering dominates the radiative effects in the wavelength range below $1 \ \mathrm{\mu m}$, where the maximum of the incident stellar spectra is located. With a scattering albedo of nearly $\omega_\lambda \approx 1$, absorption is negligible in this wavelength range. Thus the clouds will mostly scatter the incident light at short wavelengths. However, due to high asymmetry parameters ($g_\lambda \approx 0.8$) much of the stellar radiation will be scattered in the forward direction, i.e. it will still reach the planetary surface, and only a small part is reflected back to space.
In the longer wavelengths range of the outgoing thermal radiation around $10 \ \mathrm{\mu m}$, the radiative effects are much more complex, because the optical properties, especially the absorption and scattering optical depths, show large variations.
The different effects for this case are explained in detail in Sect \ref{secSurfaceTemperature}.

\section{Radiative-convective climate model}
\label{secClimateModel}

\subsection{Basic assumptions}

For the atmospheric model calculations we use a one-dimensional radiative-convective climate model, which is based on the model developed and described by \citet{Kasting1984} and \citet{Pavlov00}. In their atmospheric model the impact of clouds is not explicitly treated. The effects of clouds are only taken into account indirectly by adjusting the value of the planetary surface albedo to mimic the influence of clouds in the troposphere. Using the developed cloud model description (Sect. \ref{secCloudDescription}) we include the climatic effects of multi-layered clouds directly into the climate model to determine for instance the radiative feedbacks of clouds on the surface temperature. Chemical feedbacks of clouds are not included yet. The atmospheric profiles of the major chemical species obtained with a detailed photochemical model \citep{Grenfell07} representing the modern Earth atmosphere are used.
Since $\mathrm{N_2}$, $\mathrm{O_2}$, and $\mathrm{CO_2}$ are well mixed within the atmosphere, their atmospheric profiles are given by isoprofiles with mixing ratios of $78\%$, $21\%$ and $0.0355\%$ respectively. The profiles of $\mathrm{CH_4}$, $\mathrm{O_3}$ and $\mathrm{N_2 O}$ have been derived from the photochemical model for the modern Earth. This atmospheric composition is assumed for all calculations, thereby neglecting any change of processes influencing the chemical composition of the planetary atmospheres, like different $\mathrm{CO_2}$ levels due to changes in silicate weathering, for example.

\begin{figure}
  \centering
  \resizebox{\hsize}{!}{\includegraphics{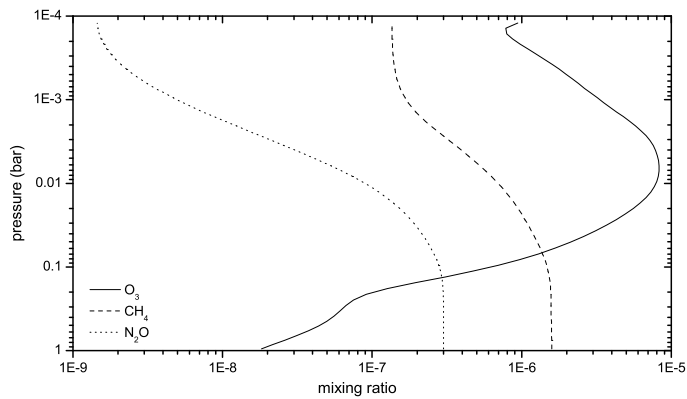}}
  \caption{Atmospheric chemical profiles of $\mathrm{O_3}$ (solid line), $\mathrm{CH_4}$ (dashed line) and $\mathrm{N_2 O}$ (dotted line). The profiles have been derived from a radiative-convective climate model coupled with a photochemistry model representing the modern Earth atmosphere (see \citet{Grenfell07}).}
  \label{chemprof}
\end{figure}

For the relative humidity in the troposphere, from which the water profile is calculated, the empirical relative humidity distribution of \citet{Manabe67} is used. Within the troposphere the temperature is assumed to follow a moist adiabate, otherwise the temperature is calculated by the condition of radiative equilibrium.
The measured Earth surface albedo of $0.13$ is applied, taken from globally averaged satellite measurements of the ISCCP. The surface emission is treated as black-body radiation determined by the surface temperature.

\subsection{Radiative transfer}
\label{subSec_RadiativeTransfer}

The radiative transfer within the climate model is split into two wavelength regimes: the stellar part, dealing with the wavelength range of the incident stellar radiation and the infrared part for the treatment of the thermal radiation wavelength range.

The radiative transfer in the stellar part consists of 38 spectral intervals between $0.238 \ \mathrm{\mu m}$ and $4.55 \ \mathrm{\mu m}$. The plane-parallel equation of radiative transfer is solved by a $\delta$-two-stream quadrature method \citep{Toon89}. Gaseous absorption of $\mathrm{O_3}$, $\mathrm{O_2}$, $\mathrm{H_2 O}$, $\mathrm{CO_2}$, and $\mathrm{CH_4}$ is treated with four-term correlated-k coefficients (cf. \citet{Segura03}). For $\mathrm{O_2}$, $\mathrm{CO_2}$ and $\mathrm{N_2}$, Rayleigh scattering is also considered. 

For the IR part a hemispheric-mean two-stream method is used, since the $\delta$-two-stream method used in the stellar part cannot be used for thermal radiation due to numerical inaccuracies \citep{Toon89}. The IR radiative transfer uses 16 spectral intervals between $3 \ \mathrm{\mu m}$ and $1000 \ \mathrm{\mu m}$. 

Gaseous absorption is treated with the correlated-k method incorporated within the rapid radiative transfer model (RRTM) developed by \citet{Mlawer97}. Considered species in the RRTM for gaseous absorption in the IR wavelength range are $\mathrm{O_3}$, $\mathrm{N_2 O}$, $\mathrm{H_2 O}$, $\mathrm{CO_2}$, and $\mathrm{CH_4}$. Scattering is neglected for gaseous species at these wavelengths. The same approach has been used for instance by \citet{Segura03}.

Note that the k-distributions used in the RRTM are only tabulated over a limited pressure and temperature range. In particular the temperature range is limited to within $\pm 30 \ \mathrm{K}$ of an Earth mid-latitude summer temperature profile (see \citet{Mlawer97} for details). Beyond the tabulated range extrapolation is used to derive the k-coefficients, which might cause inaccuracies in the upper parts of the atmospheric temperature profiles, especially for situations strongly deviating from (mean) Earth conditions (see also \citet{Segura03,Segura05} and \citet{vonParis2008}).

\subsection{Incident stellar spectra for typical M, G, K, and F-type stars}
\label{secSpectra}

Atmospheric calculations of Earth-like planets around different main sequence dwarf stars, M, F, G, and K-type stars, been have published by e.g. \citet{Kasting1993} or \citet{Segura03,Segura05}. In order to be comparable we investigated the same stars in this study as representatives for the different stellar types. The sample of stars used for the calculations are the F-dwarf $\sigma$ Bootis (HD 128167) as a typical F-type star, the Sun as the G-type star, the young active K-type star $\epsilon \ \mathrm{Eridani}$ (HD 22049), and the M-type dwarf star AD Leo (GJ 388). The basic stellar parameters (stellar type, effective temperature $\mathrm{T_{eff}}$, distance of the considered star to the Sun $d$, and distances from the planets to their host stars $a$, for which the stellar flux matches the solar constant) with corresponding references are shown in Table \ref{tabSpectraDistances}.

\begin{table}
  \caption[]{Properties of the different central stars.}
  \label{tabSpectraDistances}
  \centering
  \begin{tabular}{l l l l l l}
    \hline
    \noalign{\smallskip}
    Name & Type & $\mathrm{T_{eff}}$ (K) & $d$ (pc) & $a$  (AU)\\
    \hline
    \noalign{\smallskip}
    $\sigma \ \mathrm{Bootis}^{\mathrm{a,b}}$ & F2V & 6722 & 15.5 & 1.89\\
    Sun$^\mathrm{c}$ & G2V & 5777 & 0 & 1.0\\
    $\epsilon \ \mathrm{Eridani}^{\mathrm{a,d}}$ & K2V & 5072 & 3.2 & 0.61 & \\
    AD Leo$^{\mathrm{e,f}}$ & M4.5V & 3400 & 4.9 & 0.15\\
    \hline
    \noalign{\smallskip}
  \end{tabular}
  \begin{list}{}{}
    \item[$^{\mathrm{a}}$] \citet{Habing2001}; $^{\mathrm{b}} \ $ \citet{Cenarro2007}; $^{\mathrm{c}} \ $ \citet{CoxAstroQuant}
    \item[$^{\mathrm{d}}$] \citet{Santos2004}; $^{\mathrm{e}} \ $ \citet{Leggett1996}; $^{\mathrm{f}} \ $ \citet{Segura05}
  \end{list}
\end{table}

High resolution spectra for the M-dwarf and the F-type star were taken from \citet{Segura03,Segura05}. According to \citet{Segura03} the F-type star spectrum is a composition from IUE satellite spectra of $\sigma$ Bootis between $115 \ \mathrm{nm}$ and $335 \ \mathrm{nm}$ and a synthetic spectrum derived from the well-established stellar atmosphere model of Kurucz \citep{Kurucz79,Buser92}. More details on the F-star spectrum can be found in \citet{Segura03}. The M-type star high resolution spectrum is a composite of IUE satellite data between $115.1 \ \mathrm{nm}$ and $334.9 \ \mathrm{nm}$ and an optical spectrum from \citet{Pettersen1989} between $335.5 \ \mathrm{nm}$ and $900 \ \mathrm{nm}$. It was extended to $2409 \ \mathrm{nm}$ in the near infrared using spectra by \citet{Leggett1996} and a synthetic photospheric spectrum from the stellar atmosphere model NextGen beyond $2410 \ \mathrm{nm}$. For further details on the M-star spectrum we refer to \citet{Segura05}. 

The K-type star spectrum is composed of IUE satellite data from $\epsilon$ Eridani between $115 \ \mathrm{nm}$ and $355 \ \mathrm{nm}$ and a synthetic NextGen spectrum, taken from the grid of stellar atmosphere models of France Allard (http://perso.ens-lyon.fr/france.allard/). The high resolution spectrum of the Sun is taken from \citet{Gueymard2004}. This updated compilation of stellar spectra has been used in this study.

\begin{figure}
  \centering
 \resizebox{\hsize}{!}{\includegraphics{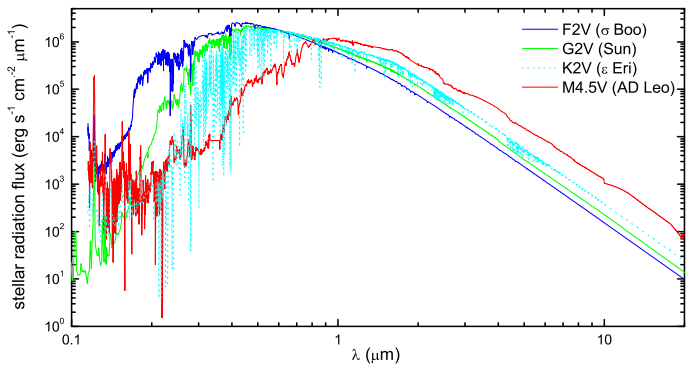}}
  \caption{Incident stellar spectra for different central stars. Each radiation flux is scaled to the distance where the total energy input at the top of the planetary atmosphere equals the solar constant.}
  \label{stellarspectra}
\end{figure}

A total integrated solar radiation flux of $1366 \ \mathrm{W m^{-2}}$ has been derived by integration of the high resolution spectrum of \citet{Gueymard2004} from $0.5 \ \mathrm{nm}$ to $1000 \ \mathrm{\mu m}$. In contrast to the approach of \citet{Segura03, Segura05}, the distances of the four representative Earth-like extrasolar planets to their respective central stars have been determined in this work in a way that the integrated incident stellar flux equals this solar constant. The corresponding orbital distances of the planets are shown in Table \ref{tabSpectraDistances}, and the stellar spectra of all four central stars incident at the top of the planetary atmospheres are shown in Fig. \ref{stellarspectra}.

Furthermore, all four spectra have been binned to obtain the integrated radiation fluxes required for the $38$ spectral intervals from $0.238 \ \mathrm{\mu m}$ to $4.55 \ \mathrm{\mu m}$ used in the radiative transfer (cf. Sect. \ref{subSec_RadiativeTransfer}).

\subsection{Cloud-climate scheme}

The climate model used in this work allows for a multi-layered cloud structure: two different cloud layers and the possibility of partial overlap between both layers.

The optical properties of the clouds are calculated according to the description in Sect. \ref{subMieTheory} and the results summarised in Fig. \ref{opticaldepth}. The optical properties are assumed to be the same for all model atmospheres, i.e. no feedback of the atmospheric properties (e.g. different temperature profiles) onto the clouds is considered here. The altitude of both cloud layers is iteratively adjusted to match the measured pressure values shown in Table \ref{tabWarren07}. 

To account for the radiative effects of clouds, their optical properties (optical depths, asymmetry parameter, and scattering albedo) have been introduced into both parts of the radiative transfer scheme. In order to account for different amounts of coverages and their partial overlap of multi-layered clouds in our model we developed a flux-averaging procedure. For every distinct cloud configuration $i$ (e.g. a low-level or a high-level cloud layer, partial overlapping clouds etc.) and the clear sky case ($\mathrm{cs}$) the radiative transfer is solved separately. The mean radiative flux $F_\lambda$ is then determined by averaging all separately calculated fluxes $F_{\lambda,i}$ weighted with the corresponding cloud coverage $x_i$

\begin{equation}
  F_\lambda = (1 - x) \cdot F_{\lambda,\mathrm{cs}} + \sum_{i=1}^I F_{\lambda,i} x_i
  \quad  \mathrm{with} \quad
  x = \sum_{i=1}^I x_i \leq 1 \ .
\end{equation}

\section{Earth reference model}
\label{secEarthReference}

In order to verify the applicability of our cloud description we first performed model calculations for the modern Earth atmosphere. ISCCP measurements report a global mean Earth surface temperature value of $288.4 \ \mathrm{K}$, which we considered as reference below. Calculations for the clear sky case and for the measured Earth mean cloud cover (39.5 \% low-level and 15 \% high-level cloud cover) have been done also considering a partial overlap of $\approx 7 \%$ between the two cloud layers. The calculated values for the surface temperatures, the temperatures at the positions of the cloud layers, and the Bond albedos are summarised in Table \ref{tabEarthReference} and compared to measured values taken from ISCCP data.

\begin{table*}
  \caption[]{Summary of the different Earth reference models in comparison to measured values.}
  \label{tabEarthReference}
  \centering
  \begin{tabular}{l c c c c}
    \hline
    \noalign{\smallskip}
    Value & Clear sky & Earth mean cloud cover & including partial overlap & measured Earth values\\
    \hline
    \noalign{\smallskip}
    Surface temperature & $293.3 \ \mathrm{K}$ & $287.7 \ \mathrm{K}$ & $288.4 \ \mathrm{K}$ & $288.4 \ \mathrm{K}$ \\
    Low-level cloud temperature & $287.3 \ \mathrm{K}$ & $280.7 \ \mathrm{K}$ & $281.4 \ \mathrm{K}$ & $281.1 \ \mathrm{K}$ \\
    High-level cloud temperature & $229.2 \ \mathrm{K}$ & $220.1 \ \mathrm{K}$ & $220.8 \ \mathrm{K}$ & $227.5 \ \mathrm{K}$ \\
    Bond albedo & $0.15$ & $0.27$ & $0.26$ & $0.3$ \\
    \hline
    \noalign{\smallskip}
  \end{tabular}
\end{table*}

\subsection{Surface temperatures} 
For the clear sky case, the calculated surface temperature is about $5 \ \mathrm{K}$ too high. Using the global mean cloud coverages of $39.5 \%$ for the high-level cloud and $15 \%$ for the high-level cloud (see Sect. \ref{secCloudDescription} and Table \ref{tabWarren07}), the resulting surface temperature ($287.7 \ \mathrm{K}$) agrees much better with the observed value. As we pointed out in Sect. \ref{secCloudDescription} these measured cloud coverages already include a certain amount of overlap between the cloud layers, but this value was not specified by the measurements. We determined a partial overlap of about $7 \%$ required to reach a surface temperature of $288.4 \ \mathrm{K}$.

\begin{figure}
  \centering
  \resizebox{\hsize}{!}{\includegraphics{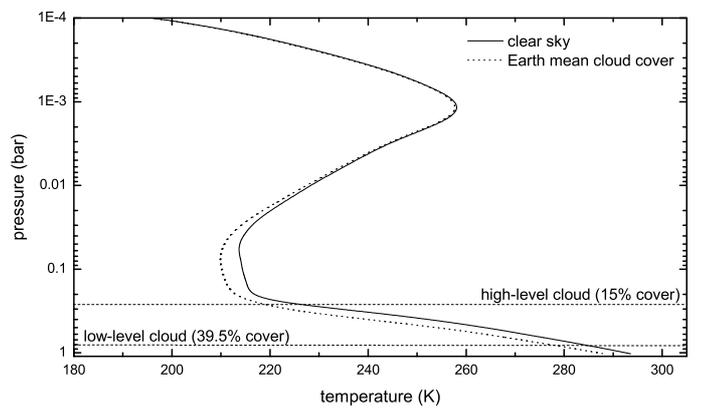}}
  \caption{Temperature-pressure profiles of the Earth reference models. The solid line represents the clear sky case, the dotted line the model with the mean Earth cloud cover. The horizontal dashed lines represent the position of both cloud layers.}
  \label{earth_temp_profile}
\end{figure}

The temperature pressure profile of the clear sky case and of the cloudy case (without overlap) are shown in Fig. \ref{earth_temp_profile}. The positions of the two cloud layers are denoted by the two dashed lines. The profile of the model including partial overlap is not shown because it shows no noticeable difference compared to the non-overlapping case. 
The profiles again indicate that the clear sky calculation results in too high temperatures, while the cloudy case resembles mean Earth conditions. The presence of clouds changes the temperature profiles up through the troposphere, while the temperatures in the upper atmosphere are almost unaffected as expected. 

The temperatures of the low-level clouds also agree very well with the measured values. In contrast to that, the high-level cloud temperatures show a difference of about $6.7 \ \mathrm{K}$ in comparison to the values published by \citet{Rossow99a} (see Table \ref{tabWarren07}). Still, as was pointed out in their work, the cloud temperatures were derived neglecting cloud IR scattering. This resulted in an overestimate of their high-level cloud temperatures by a few degrees so that the actual deviations of our calculated cloud temperatures are much smaller than $7 \ \mathrm{K}$. Because the optical depth and the coverage of the low-level clouds is larger than that of high-level clouds, clouds in the present Earth atmosphere have a net cooling effect as confirmed by our model findings.

\subsection {Bond albedo}

The calculated value of 0.15 for the planetary Bond albedo (see Table \ref{tabEarthReference}) is much too small in the clear sky case compared with the observed Earth value of about $0.3$. With clouds included in the model, the resulting Bond albedo of 0.27 (0.26 in case of overlapping clouds) agrees much better with the observed value. The remaining small difference can be explained by our neglect of the mid-level and cumulonimbus clouds, which would also contribute to the Bond albedo.

\subsection{Radiative flux profiles}

\begin{figure}
  \centering
  \resizebox{\hsize}{!}{\includegraphics{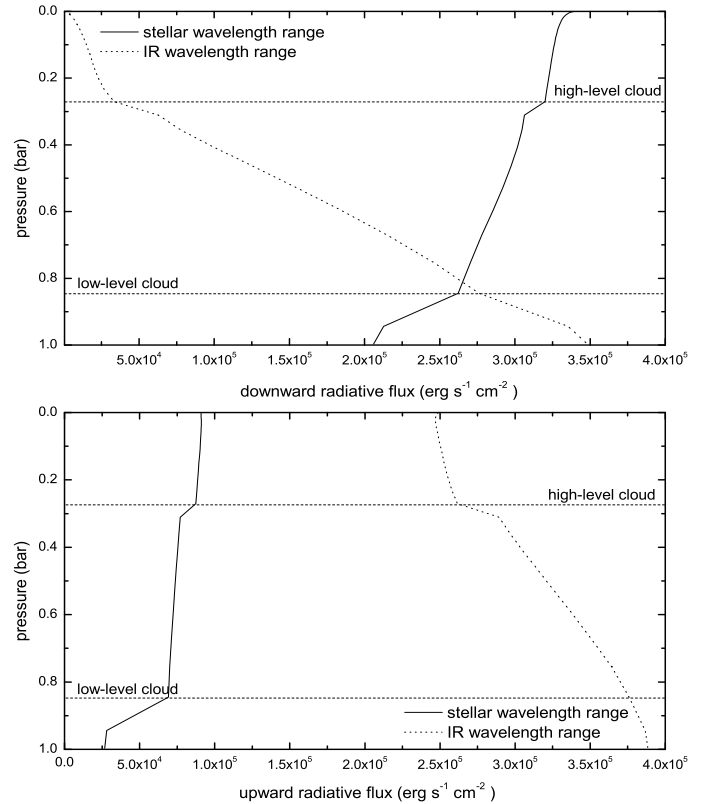}}
  \caption{Radiative flux-pressure profiles of the Earth reference calculation including clouds. The upper diagram shows the downward radiative flux in the stellar (solid line) and the IR wavelength range (dotted line), the lower diagram the upward radiative flux. The position of the two cloud layers is denoted by the horizontal dashed lines.}
  \label{earth_rad_flux}
\end{figure}
\begin{figure*}
  \centering
  \resizebox{\hsize}{!}{\includegraphics{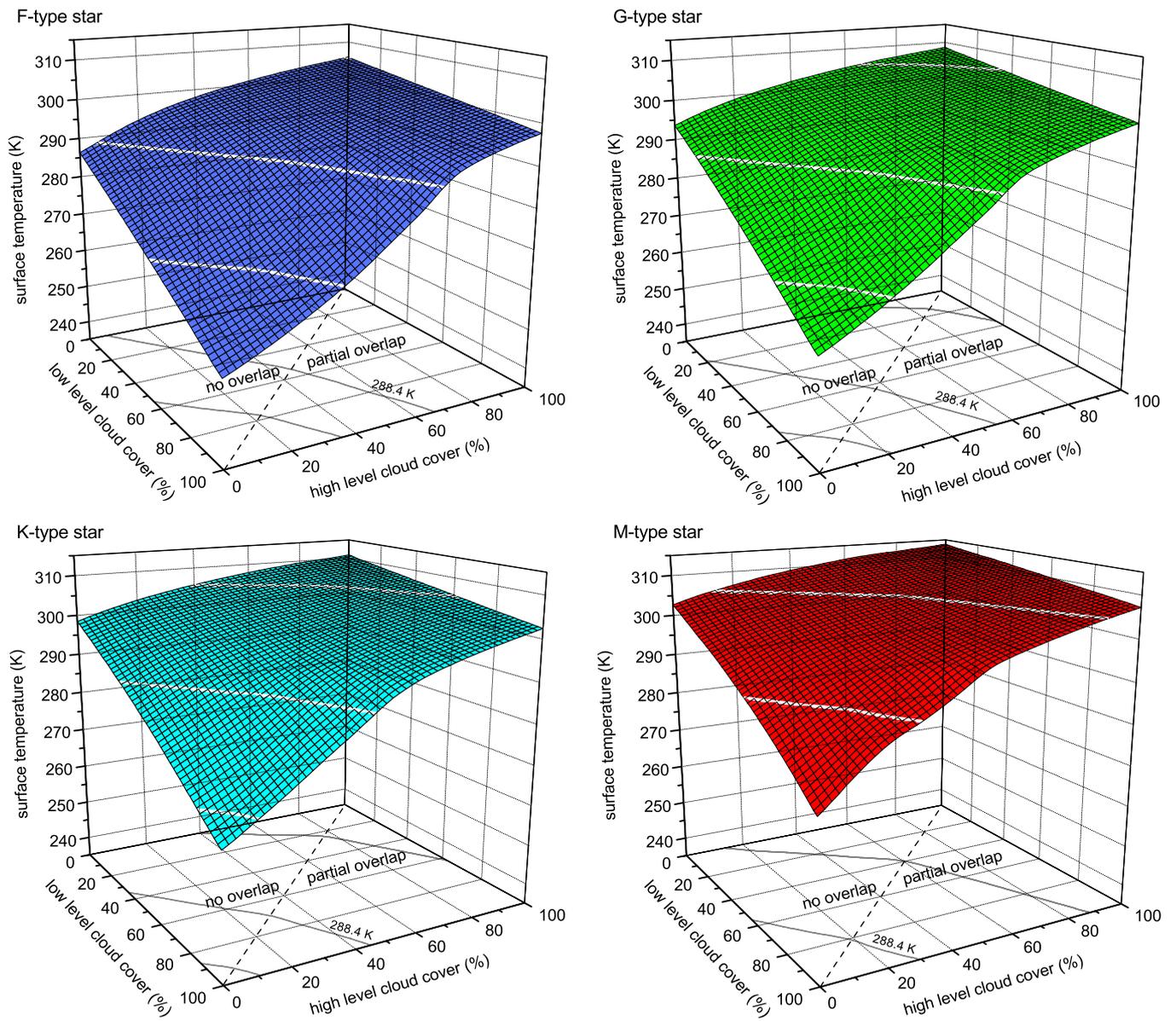}}
  \caption{Surface temperatures of planets around the four different stars considered as a function of the cloud coverages of high and low-level clouds. Upper left diagram: F-type star, upper right diagram: G-type star, lower left diagram: K-type star, lower right diagram: M-type star. The solid lines on the contour surface and in the x-y-plane denote the physical limits of the cloud parametrisations. The parameters, for which a mean Earth surface temperature ($288.4 \ \mathrm{K}$) is obtained are also marked by solid lines (see text for details). Each contour square side represents a change of $2.5\%$ in cloud coverage. The dashed line in the x-y-plane separates the parameter regions in which partial overlap of the two cloud layers occurs or not.}
  \label{surface_temp}
\end{figure*}

For a better understanding of the cloud radiative forcing, the upward and downward radiation flux-pressure profiles in the stellar and IR wavelengths range are shown in Fig. \ref{earth_rad_flux}.

\begin{figure*}
  \centering
  \resizebox{\hsize}{!}{\includegraphics{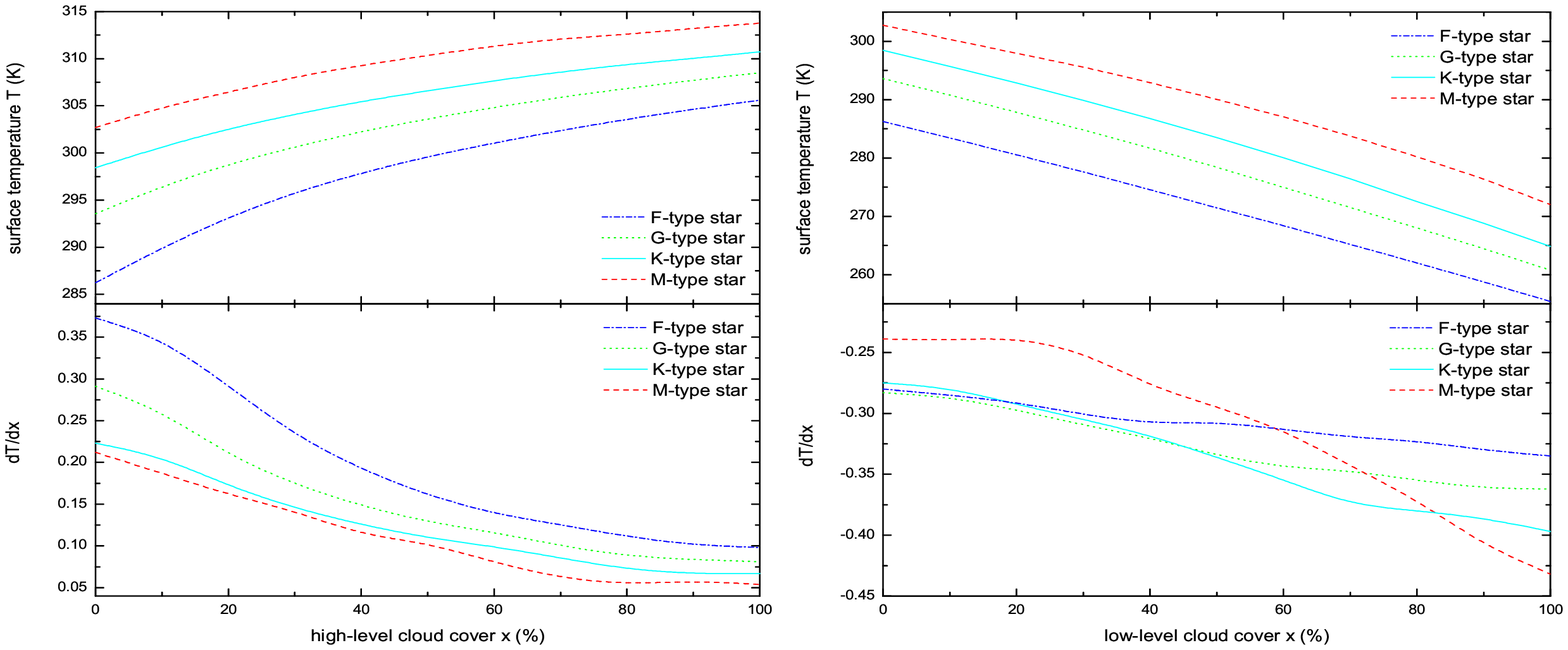}}
  \caption{Basic effect of each cloud type on the surface temperatures for the four different central stars. The upper diagrams show the surface temperature as a function of the cloud coverage of the high-level cloud (left diagram) and low-level cloud (right diagram). The lower diagram shows the derivative of the surface temperature with respect to the cloud coverage, i.e. the change of surface temperature with cloud coverage.}
  \label{cloud_basic}
\end{figure*}

The downward stellar flux clearly indicates the albedo effect of both cloud layers. Due to its larger optical depth and bigger coverage, the albedo effect of the low-level cloud is much more pronounced than that of the high-level cloud. The upward infrared flux shows the blocking of the thermal radiation by the different cloud layers, i.e. the resulting greenhouse effect. The low-level cloud has almost no effect on the outgoing IR radiation and therefore exhibits no noticeable greenhouse effect. This yields a net albedo effect (see the steps in the flux profiles in Fig. \ref{earth_rad_flux}), which leads to a cooling of the lower atmosphere and also of the planetary surface. The high-level cloud on the other hand traps more IR radiation in the lower atmosphere than it blocks the incident solar radiation. Thus, for the high-level clouds the greenhouse effect exceeds their albedo effect, which leads to a net heating in the lower atmosphere and an increase in the surface temperature (see also Fig. \ref{cloud_basic}).

To summarise, our parametrised cloud model is able to reproduce the mean Earth conditions, using measured cloud properties, cloud coverages, and the Earth global mean surface albedo. The Earth reference model can reproduce the mean Earth surface temperature and the Earth Bond albedo very well. The temperatures at the cloud positions also agree favourably with measured values.

To mimic the climatic effects of clouds it is a common approach to adjust the planetary surface albedo in one-dimensional clear sky calculations (cf. e.g. \citet{Segura03}). While this approach can reproduce the correct surface temperature, these models have the shortcoming, amongst other things, that they are unable to reproduce the correct planetary Bond albedo. The parametrised cloud model, though, is able to reproduce both parameters.

\section{Earth-like planetary atmospheres}
\label{secEarthLikeAtmos}

\begin{figure*}
  \centering
  \resizebox{\hsize}{!}{\includegraphics{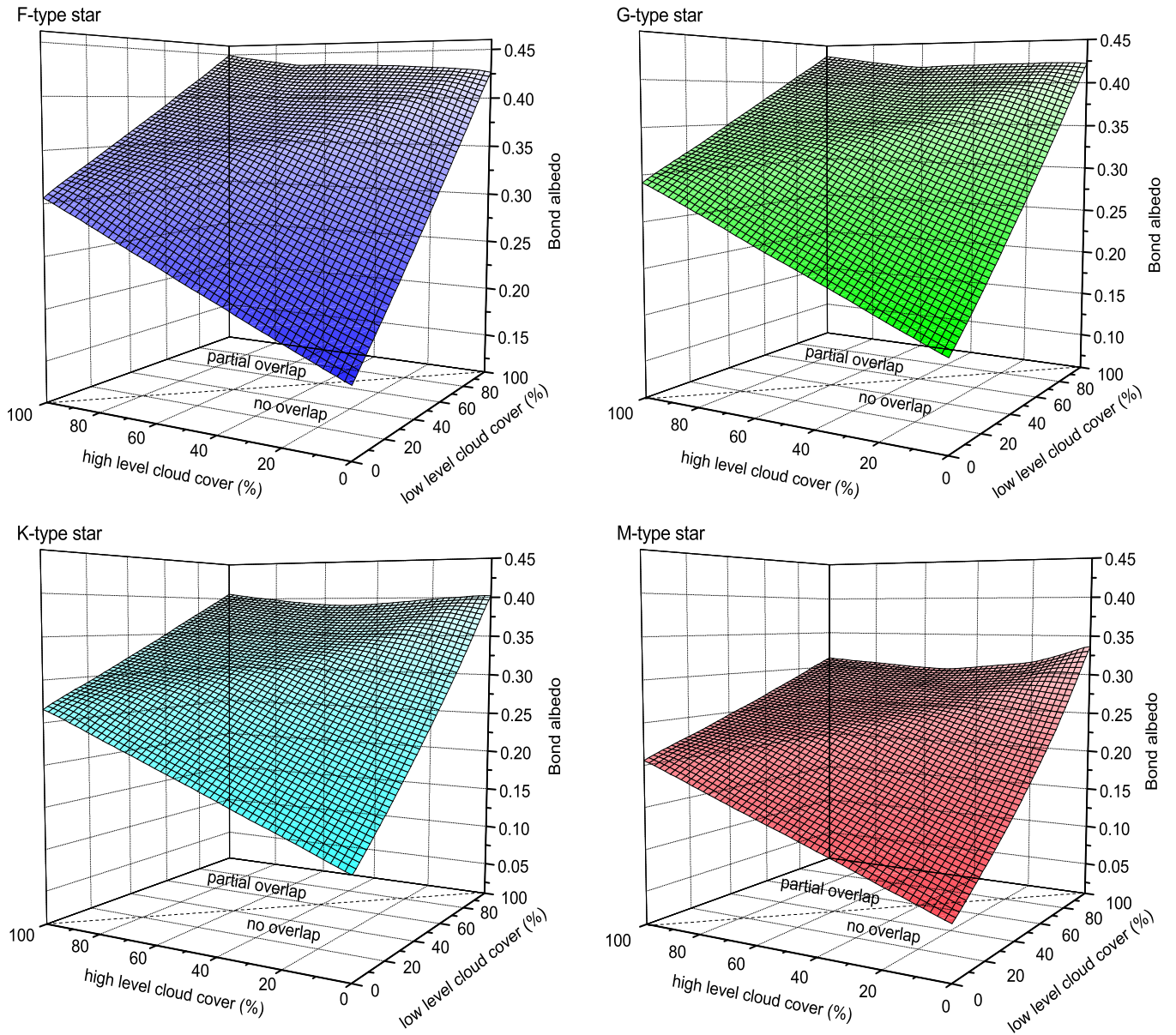}}
  \caption{Planetary Bond albedo of planets around the four different stars as a function of the cloud coverages of high and low-level clouds. Upper left diagram: F-type star, upper right diagram: G-type star, lower left diagram K-type star, lower right diagram M-type star. Each contour square size represents a change of $2.5\%$ in the cloud coverage. The dashed line in the x-y-plane separates the parameter regions in which partial overlap of the two cloud layers occurs or not.}
  \label{bond_albedo}
\end{figure*}

In this section we extend our studies of the climatic effects of clouds to Earth-like extrasolar planetary atmospheres at orbital distances for which the stellar energy input at the top of the atmosphere equals the solar constant. The cloud parametrisation used is the same as for the Earth, but the cloud coverages of the high and low-level cloud layers are varied. To limit the number of free parameters the cloud overlap is calculated from the cloud coverages. As long as the sum of the coverages of both cloud layers is $\leq 100 \%$, no overlap occurs. If the sum of the two cloud coverages exceeds $100 \%$, the cloud layers are partially overlapping with the smallest amount of overlap possible. As an example, a low-level cloud cover of $20 \%$ and high-level cloud cover of $80 \%$ results in no overlap, while for $60 \%$ and $80 \%$ coverages, the clouds are overlapping by $40 \%$. Thus the resulting two dimensional parameter space, spanned by the two cloud coverages, is divided into two regimes of partially overlapping clouds and clouds without overlap (see Figs. \ref{surface_temp} and \ref{bond_albedo}). 

In order to quantify the climatic effects of clouds in Earth-like planetary atmospheres calculations have been carried out over the whole range of the two-dimensional parameter space of possible cloud cover combinations. For each stellar type more than $150$ atmospheric models have been evaluated. The resulting surface temperatures and Bond albedos are presented in the next subsections. First implications of the position of the habitable zones of the four different central stars considered are discussed at the end of this section.

\subsection{Surface temperatures}
\label{secSurfaceTemperature}

Figure \ref{surface_temp} shows the surface temperatures as a function of the cloud coverages for the four different central stars.
The various lines shown on the contour surfaces indicate different regions of physical conditions. The lowest line marks the validity range for the low-level cloud. For parameters below the corresponding line in the x-y-plane, the temperature of the low-level cloud drops below $260 \ \mathrm{K}$, which is the lower temperature limit for the approach (see Sect. \ref{secCloudDescription}), i.e. it represents the freezing limit for water droplet clouds. The uppermost line on the other hand indicates the validity range of the high-level clouds. Above that line the high-level cloud temperature is higher than $260 \ \mathrm{K}$, where the ice cloud would li\-quefy. The middle line denotes the parameters for which a mean Earth surface temperature of $288.4 \ \mathrm{K}$ occurs. It should be noted that all three lines become important only for the K and G-type stars. Due to the low resulting temperatures for the F-type star, the limit for the high-level cloud is not reached, while for the M-type star the temperature of the low-level cloud always stays above $260 \ \mathrm{K}$.

Using the measured Earth surface albedo, results imply that none of the four clear sky model atmospheres achieves surface temperatures equal to the Earth mean value of $288.4 \ \mathrm{K}$ \footnote{This is of course the reason why the surface albedo has to be adjusted to mimic the effect of clouds in clear sky model calculations for Earth.}. Even if the incident stellar energy is the same for all types of stars and equals the solar constant, the resulting clear sky surface temperatures are obviously different (see Figs. \ref{surface_temp}, \ref{cloud_basic}), which makes the influence of the wavelength dependence of the incident radiation quite clear, although no clouds are present and only the wavelength dependent opacities of the gas are causing the difference in these cases. The surface temperature of a cloudless Earth-like planet orbiting the F-type star, for example, is lower than the measured mean Earth surface temperature, while the other surface temperatures are higher than $288.4 \ \mathrm{K}$. This is especially the case for the M-type star, which has substantially more radiation flux in the visible and IR wavelength regions than the other three kinds of stars (cf. Fig. \ref{stellarspectra}). Consequently clouds are required to reach $288.4 \ \mathrm{K}$ in all investigated situations. Whereas the cloud layers have to produce a net heating effect in the case of the F-type star, the albedo effect of clouds has to be stronger than the greenhouse effect for the other types of stars if mean Earth surface temperatures are to be achieved under such conditions. 

In order to illustrate the basic climatic effect, the results of single cloud layer calculations are shown in Fig. \ref{cloud_basic}, which summarises the corresponding slices through the 3D-temperature diagrams of Fig. \ref{surface_temp}. As a rule the surface temperature decreases with increasing low-level cloud coverage, i.e. water droplet clouds exhibit a net albedo effect for all stellar spectra. The maximum temperature decrease of about $30 \ \mathrm{K}$ at full cloud cover is about the same for all types of central stars. The change of the surface temperature with increasing low-level cloud coverage is, however, not uniform. The most significant difference appears for the M-type star, while the for the F-type star there is almost no change in the temperature (see lower right panel of Fig. \ref{cloud_basic}). This effect is caused by different absorption properties of the clouds at different wavelengths in conjunction with the incident stellar spectra. The M-star spectrum (Fig. \ref{stellarspectra}) is more extended into the IR where the frequency dependent spectral energy can be partly absorbed by the low-level water cloud (cf. Fig. \ref{opticaldepth}). In contrast to the M-type star, the spectrum of the F-type star has its maximum flux at a wavelength range around $0.5 \ \mathrm{\mu m}$, where the water cloud has an almost constant scattering albedo of about $1$ (see Fig. \ref{opticaldepth}), which means that scattering dominates the radiative transfer, and the remaining absorption has only a minor influence on the temperature.

\begin{figure}
  \centering
  \resizebox{\hsize}{!}{\includegraphics{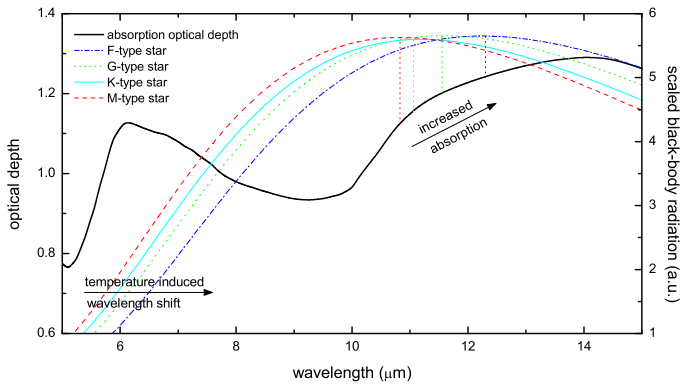}}
  \caption{Illustration of the efficiency of the greenhouse effect at different atmospheric temperatures of the layer below the high-level cloud. The solid line represents the absorption optical depth of the high-level ice cloud. The black-body radiation fluxes for planets around the different central stars (F-type star: dashed-dotted, G-type star: dotted, K-type star: solid, M-type star: dashed) were scaled for comparison. The temperatures were calculated with $50 \%$ high-level cloud cover. The vertical dotted lines denotes the positions of the maxima of the black-body radiation fluxes.}
  \label{hlc_heating}
\end{figure}

Due to their net greenhouse effect the high-level ice clouds cause an increase in the surface temperature with increasing cloud cover in all cases (see Fig. \ref{cloud_basic}). The maximum of the temperature increase caused by the greenhouse effect, however, depends on the properties of the central star considered. The albedo effect depends directly on the incident stellar spectrum, but the greenhouse effect is determined by the thermal emission of the lower atmosphere, which itself is an indirect consequence of the stellar radiation. For example an Earth-like planet around the M-type star has a maximum greenhouse effect of about $11 \ \mathrm{K}$ at $100 \%$ high-level cloud cover compared to the clear sky case. In the case of the F-type star the greenhouse effect results in an increase of surface temperature causing a temperature alteration of about $20 \ \mathrm{K}$. In principle the greenhouse effect becomes smaller for larger temperatures in the lower atmosphere (Fig. \ref{cloud_basic}, left panel). The effectiveness of the greenhouse effect depends directly on the absorption characteristics of the high-level cloud in the thermal wavelength range in conjunction with the temperature dependent infrared emissions of the lower atmosphere.

To illustrate the absorption characteristics of the clouds as a function of different atmospheric temperatures, we assume that the transmission of the atmospheric layer below the cloud layer can be represented by black-body radiation with the corresponding atmospheric temperatures. Figure \ref{hlc_heating} shows the absorption optical depth of a high-level ice cloud with $50 \%$ coverage in comparison to black-body radiation fluxes of different atmospheric temperatures derived from model calculations for the four different central stars. Compared with the other Earth-like cases, the planet around the F-type star has the lowest atmospheric temperatures, which results in the biggest shift of the black-body radiation to longer wavelengths. Since the absorption at the position of the maximum of the black-body radiation is more than $12 \%$ smaller for the M-star than for the F-star, the greenhouse effect is much more significant in the latter case. Thus one obtains a temperature-induced wavelength shift for the absorbable radiation, which results in increasing absorption and a more effective greenhouse effect. As the temperature in the lower atmosphere increases with larger high-level cloud cover, the maximum of the black-body radiation is more and more shifted to smaller wavelengths which in turn decreases the efficiency of the greenhouse effect. The amount of the greenhouse effect is in that way self-limited. This can clearly be seen in Fig. \ref{cloud_basic} (left lower panel) which shows the flattening of the temperature gradients at high ice cloud coverages.

The climatic influence of clouds in Earth-like planetary atmospheres becomes much more complex for a multi-layered situation, where the greenhouse effect and the albedo effect of the high and low-level clouds are interacting non-linearly, resulting in the 3D-temperature planes displayed in Fig. \ref{surface_temp}. The efficiency of the ice cloud greenhouse effect, for example, is much more enhanced if a low-level cloud is present below the high-level cloud. The albedo effect of the low-level cloud is compensated by the ice cloud greenhouse effect. Therefore the resulting temperature difference for increasing high-level cloud coverage is much more pronounced at $100 \%$ low-level cloud cover than for a single ice cloud layer. This can clearly be inferred from Fig. \ref{surface_temp} for all four considered scenarios. 

\begin{figure}
  \centering
  \resizebox{\hsize}{!}{\includegraphics[]{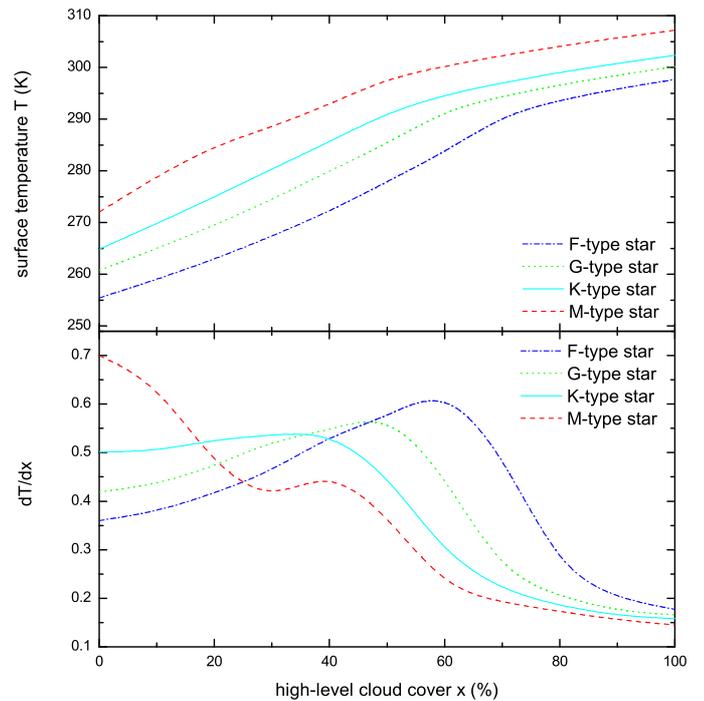}}
  \caption{Effects of multi-layered clouds on the surface temperatures for the four different central stars. The upper diagram shows the surface temperature in dependence of the high-level cloud coverage with $100\%$ low-level clouds. The lower diagram shows the derivate of the surface temperature with respect to the cloud coverage.}
  \label{cloud_enhanced_greenhouse}
\end{figure}

The enhancement of the greenhouse effect is, of course, a result of lower atmospheric temperatures, reduced due to the presence of the low-level cloud and its albedo effect. Due to the reduced temperatures the intensity maximum of the absorbable radiation is shifted to longer wavelengths, thereby increasing the efficiency of the greenhouse effect (cf. Fig. \ref{hlc_heating}). The increase of the surface temperature at \textit{full} low-level cloud cover with increasing high-level cloud coverage is shown in detail in Fig. \ref{cloud_enhanced_greenhouse}, which combines four different slices through the 3D-surface temperature diagrams of Fig. \ref{surface_temp}, respectively. Obviously, the temperature increase becomes weaker at large ice cloud coverages for all cases, as also revealed by the maxima of the corresponding temperature gradients (F-type star: $\approx 60 \%$, G-type star $\approx 50 \%$, K-type star $\approx 40 \%$, M-type star $\approx 40 \%$) shown in the lower panel of Fig. \ref{cloud_enhanced_greenhouse}. This is caused by the change of the energy transport mechanism from radiative transfer (at smaller high-level cloud coverages) to convection (for larger high-level cover) at the calculated heights of the upper cloud layers in the planetary atmospheres. The switch from convection to radiative transfer is also responsible for the second feature in the temperature slope of the M-star planet occurring at low ice cloud coverages. However, note that the highest (lowest) values of the surface temperatures affected by clouds are reached for single ice clouds (low-level clouds) with $100\%$ coverage (Fig. \ref{surface_temp}).\footnote{The applicability limitations of the cloud description should thereby be kept in mind.}
The difference between the maximum and the minimum surface temperature again depends on the characteristics of the central star. The F-type star causes the largest temperature variation ($\approx 50 \ \mathrm{K}$), whereas the temperature difference for the M-type star is the smallest ($\approx 40 \ \mathrm{K}$).

We note that the mean Earth surface temperature, which requires the presence of clouds, is not uniquely obtained by one single combination of low and high-level cloud coverages. On the contrary, several cloud cover combinations can in principle result in a surface temperature of $288.4 \ \mathrm{K}$ as indicated by the corresponding contour lines in Fig. \ref{surface_temp}. However, the range of coverage combinations for which mean Earth surface conditions can be reached differs between the stellar types from the M-type star (smallest parameter range) to the F-type star (largest parameter range). Due to its higher incident stellar flux in the visible and IR wavelength region, the M-type star requires a high cover (about $60 \%$) of low-level clouds as a minimum to achieve a cooling to $288.4 \ \mathrm{K}$. For the F-type star on the other hand only about $10 \%$ of high-level clouds are needed for the required additional heating.

\subsection{Bond albedo}

The Bond albedo is generally an important quantity to characterise the influence and properties of clouds in planetary atmospheres. Here the Bond albedo values have been obtained for the whole two-dimensional parameter space of cloud coverages. In Fig. \ref{bond_albedo} the calculated planetary Bond albedos are shown for all four central stars considered. As expected, the albedo increases regardless of the cloud type with increasing cloud cover; i.e. in absence of any cloud the Bond albedos are given by the properties of the gas only and have very low minimal values between $0.07$ for the M-type star and $0.17$ for the F-type star. Overall, the albedo values for an Earth-like planet around an M-type star are lower than for the other stellar types, whereas for the F-type star the Bond albedos are the highest. This is the result of direct absorption of the incident stellar radiation by clouds and gas. Since the clouds can absorb a larger fraction of the M-star spectrum (see Figs. \ref{opticaldepth} and \ref{stellarspectra}), less of that light is reflected back into space, resulting in a lower Bond albedo.

The low-level clouds have a larger impact upon the Bond albedos than the ice clouds because their optical depth is twice as large in the wavelength range of the stellar spectra (cf. Figs. \ref{opticaldepth} and \ref{stellarspectra}). Therefore the Bond albedos caused by low-level clouds alone are higher than the albedos of single ice cloud layers. One would expect that the Bond albedos become maximal in the case of total cloud cover of both cloud layers. This is the case for the F-, G- and K-type conditions, whereas for the M-type star the single low-level cloud layer with $100 \%$ coverage yields the highest Bond albedo. This is related to a slight anomaly which is evident for the M-type star and also (but less pronounced) for the K-type star (Fig. \ref{bond_albedo}). At total low-level cloud cover the Bond albedo starts to decrease with \textit{increasing} high-level cloud coverage. Where the ice cloud cover exceeds about $50 \%$ the albedo values increase again.

\begin{figure}
  \centering
  \resizebox{\hsize}{!}{\includegraphics{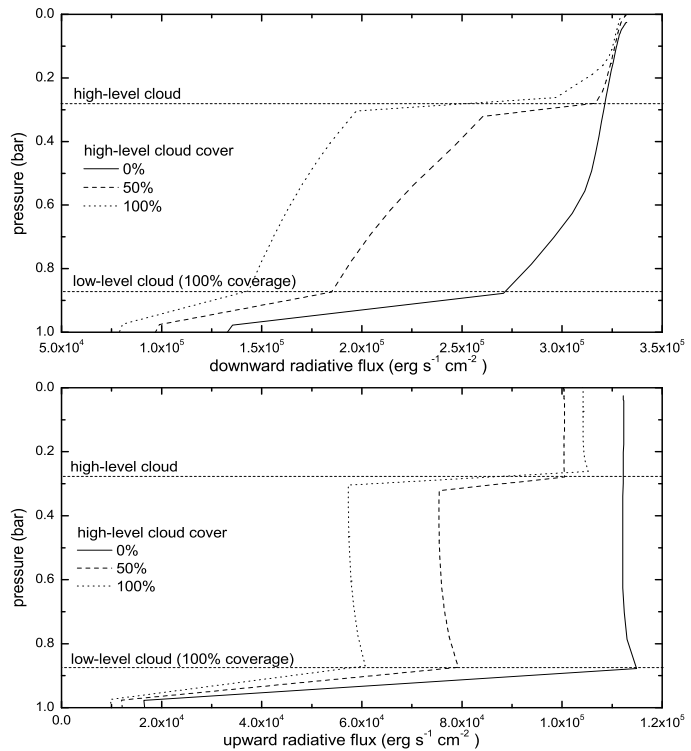}}
  \caption{Radiative flux-pressure profiles of different model calculations for the M-type star to illustrate the albedo anomaly. The calculations have been performed with a $100\%$ cover of low-level clouds and three different high-level cloud coverages. The upper diagram shows the downward radiative flux in the stellar wavelength range, the lower diagram the upward radiative flux. The position of the two cloud layers is denoted by the horizontal dashed lines.}
  \label{AlbedoAnomaly_flux}
\end{figure}

This effect can be illustrated by the radiative flux-pressure profiles for different high-level cloud coverages and full low-level cloud cover in the atmosphere of an Earth-like planet orbiting an M-star (Fig. \ref{AlbedoAnomaly_flux}). The ice cloud presence affects the downward radiative flux as expected (upper panel of Fig. \ref{AlbedoAnomaly_flux}) by enhanced absorption and scattering capability. But above the upper cloud level the upward radiation flux for full ice cloud coverage is \textit{higher} than the corresponding radiation flux for only $50 \%$ cover (see lower diagram of Fig. \ref{AlbedoAnomaly_flux}). This can be explained by increased gas absorption of the incident stellar radiation for these specific atmospheric conditions.
This effect deserves further investigation especially in view of improved gas opacities.

\subsection{Habitable zone positions influenced by $\mathrm{H_2 O}$ clouds - first implications}

The potential habitability of terrestrial planets depends on their surface conditions, especially on the surface temperatures. Usually the possible existence of liquid water on the planetary surface is considered as an indication for habitable conditions. Below we refer to the measured mean Earth surface temperature of $288.4 \ \mathrm{K}$ as characteristic for the positions of habitable zones around different kinds of central stars.

\begin{figure}
  \centering
  \resizebox{\hsize}{!}{\includegraphics{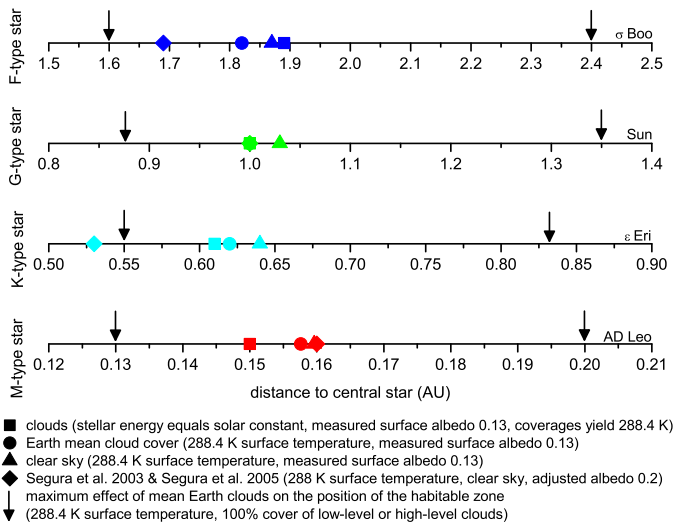}}
  \caption{Positions of the habitable zone around different types of central stars affected by clouds. Squares mark the distances at which the incident stellar flux matches the solar constant. Circles indicate the distances for which a mean Earth surface temperature is achieved using the measured Earth cloud cover. The triangles show the positions for planets with clear sky atmospheres and a mean Earth surface temperature. The corresponding distances derived by \citet{Segura03, Segura05} using a clear sky model with adjusted surface albedo are marked by diamonds. The maximum effect of the clouds on the distances are marked by arrows.}
  \label{star_distances}
\end{figure}

As already pointed out in Sect. \ref{secSpectra} the orbital distance of the Earth-like planets to their host stars have been determined here in a way that the incident stellar energy matches the solar constant regardless of the atmospheric details (see Table \ref{tabSpectraDistances}). Using these distances, the calculated planetary surface temperatures do not comply with the mean Earth surface temperature assuming clear sky conditions as discussed in detail in Sect. \ref{secSurfaceTemperature}. Habitable conditions can yet be achieved even for these positions due to the complex climatic effects caused by $\mathrm{H_2 O}$ cloud layers as for the Earth for example. As previously described, mean Earth surface temperature conditions can in principle be reached for several cloud coverage combinations. Adjusting the stellar energy input at the top of the atmosphere by changing the distance between the central star and planet properly can, however also result in Earth-like conditions for clear sky atmospheres. The deviations between these approaches can clearly be seen in Fig. \ref{star_distances}.

Using clear sky atmospheres the orbital distances of the planets have to be increased (by about $2 \%$ for the M-type star and $3 \%$ in the case of G-type and K-type stars) to compensate for the missing net cooling effect of the $\mathrm{H_2 O}$ cloud layers and to achieve Earth-like conditions on the surface. For the F-type star, the distance needs to be decreased by about $3 \%$, because in the clear sky case the surface temperature is below $288.4 \ \mathrm{K}$. Consequently, planets with a measured mean Earth cloud cover can be located closer to the central star than planets with clear sky atmospheres with the exception of the F-type star.

The positions of Earth-like planets in the habitable zones calculated by \citet{Segura03, Segura05} are also indicated in Fig. \ref{star_distances} for comparison. These distances have been derived from clear sky calculations for the Earth atmosphere with an adjusted planetary surface albedo of $0.2$ to account for the already mentioned net cooling effect of clouds\footnote{This tuning, however, is only possible if the target surface temperature is prescribed, while in case of the model including clouds the surface temperature is a result of the calculations.} and is used for all different types of central stars. \citet{Segura03, Segura05} had additionally to change the distances of these planets to their host stars to reach a temperature of $288 \ \mathrm{K}$ at the planetary surface even with the modified surface albedo. These positions deviate from the distances of planets with an Earth-like cloud cover (this work) between $17 \%$ in case of the K-type star, about $10 \%$ for the F-type star and less than $1 \%$ for the M-type star. These differences are partly the consequence of the increased surface albedo used by \citet{Segura03, Segura05}, which is $54 \%$ larger than the measured Earth mean value used here. A larger surface albedo leads to less absorbed stellar flux at the surface. Therefore the planets in the work of \citet{Segura03, Segura05} are located closer to the central star (except for the M-type star) compared to our findings including the effect of clouds. Still, it should be noted that other differences between the two models do also affect the deviation in the positions of the habitable zones. An important difference in the modelling approaches is certainly the different treatment of the atmospheric chemistry. Here Earth-like chemical profiles are used (see Fig. \ref{chemprof}), whereas \citet{Segura03, Segura05} determined the chemical composition from a photochemical model. The obvious deviation in the case of the K-type star is mainly caused by the different stellar input spectra used (cf. Sect. \ref{secSpectra}).
But apart from the well known overall effect that the positions of the habitable zone gets closer and closer to the central star by changing its spectral type from F to M (cf. \citet{Kasting1993}), the special spectral characteristics of the stars are affecting the positions of the habitable zones differently due to their complex interaction with the atmospheric cloud layers (see Fig. \ref{star_distances}).

The maximum effect of mean Earth clouds on the position of the habitable zone are denoted by arrows in Fig. \ref{star_distances}. These distances have been derived with $100 \%$ of either cloud type while still obtaining a mean Earth surface temperature of $288.4 \ \mathrm{K}$. Using low-level clouds for a maximum cooling effect, the planets can be located up to $15 \%$ closer to the central star compared to a clear sky planet. The distances can be increased by up to $35 \%$ when single high-level cloud layers are present. The distances derived by \citet{Segura03,Segura05} are in all cases within these boundaries, except for the already discussed K-type star.

It is important to note that the distances derived in this work are only indications of the position of the habitable zone, they do not indicate its boundaries\footnote{See e.g. \citet{Selsis2007} for a study on the boundaries of the habitable zone of planets around different types of central stars.}. 
The inner boundary is determined by the runaway greenhouse effect while the outer edge is determined by $\mathrm{CO_2}$ clouds (cf. \citet{Kasting1988} and \citet{Selsis2007}). Since these kinds of atmospheres are in general far from being Earth-like, our parametrised cloud description, which is based on measurements in the Earth atmosphere, is unsuitable to study the effects of clouds on the boundaries of the habitable zone. Further studies of the effects of clouds on habitable zones require therefore a cloud microphysics model to evaluate the cloud properties (e.g. optical thickness, wavelength dependent optical properties, altitude) properly, since measurements and therefore also parametrised descriptions of them are not available for these more extreme physical conditions.
A more detailed analysis of the influence of clouds on the position and extension of the habitable zone will be addressed in a forthcoming publication.

\section{Summary}
\label{secSummary}

We developed a parametrised cloud description scheme for Earth-like planetary atmospheres based on measurements of clouds in the Earth's atmosphere. The optical properties of the clouds were calculated with the Mie theory combined with an equivalent sphere approach for the ice crystals. This cloud scheme was coupled with a one-dimensional radiative convective climate model, and its applicability was tested on an Earth reference model. Model calculations were made for Earth-like planetary atmospheres of planets orbiting different types of main sequence dwarf stars: F, G, K, and M-type stars.

It was shown that the albedo effect is only weakly dependent on the incident stellar spectra because the optical properties (especially the scattering albedo) remain almost constant in the wavelength range of the maximum of the incident stellar radiation. The greenhouse effect of the high-level cloud on the other hand depends on the temperatures of the lower atmosphere, which in turn are an indirect consequence of the different types of central stars. The efficiency of the greenhouse effect increases with smaller temperatures in the lower atmosphere and decreases with higher atmospheric temperatures. As a rule the planetary Bond albedos increase with cloud cover of either cloud type. However, for the K and M-type star an anomaly was found resulting in a decreasing Bond albedo with increasing cloud cover for a certain region in the parameter space. This anomaly is caused 
by enhanced gas absorption under these specific atmospheric conditions.

Planets with Earth-like clouds in their atmospheres can be located closer to the central star or farther away compared to planets with clear sky atmospheres. The change in distance depends on the type of cloud. In general, low-level clouds result in a decrease of distance because of their albedo effect, while the high-level clouds lead to an increase in distance. 

Apart from these climatic effects, clouds also affect the planetary emission and transmission spectra, which is important concerning the detectability of spectral signatures of e.g. biomarker molecules. The investigation of the influence of clouds on these effects will be done in further studies.

\begin{acknowledgements}
  This work has been partly supported by the Forschungsallianz \textit{Planetary Evolution and Life} of the Helmholtz Gemeinschaft (HGF).
\end{acknowledgements}

\bibliographystyle{aa}
\bibliography{13491references}

\end{document}